\documentclass[aps,twocolumn]{revtex4-1}
\usepackage{amsmath,amssymb}
\usepackage{slashed}
\usepackage{graphicx}
\usepackage{bm}
\usepackage[T1]{fontenc}
\usepackage[utf8]{inputenc}
\usepackage{gauss} 
\usepackage[top=1.0in,bottom=1.0in,left=1.0in,right=1.0in]{geometry}
\usepackage[colorlinks,linkcolor=blue,citecolor=blue]{hyperref}
\usepackage{subfig}
\newcommand{\be}{\begin{equation}}
\newcommand{\ee}{\end{equation}}
\newcommand{\bea}{\begin{eqnarray}}
\newcommand{\eea}{\end{eqnarray}}
\newcommand{\ba}{\begin{eqnarray}}
\newcommand{\ea}{\end{eqnarray}}

\begin{document}

\title{Hadronic structure on the light-front  VI.\\
Generalized parton distributions of unpolarized  hadrons}

\author{Edward Shuryak}
\email{edward.shuryak@stonybrook.edu}
\affiliation{Center for Nuclear Theory, Department of Physics and Astronomy, Stony Brook University, Stony Brook, New York 11794--3800, USA}

\author{Ismail Zahed}
\email{ismail.zahed@stonybrook.edu}
\affiliation{Center for Nuclear Theory, Department of Physics and Astronomy, Stony Brook University, Stony Brook, New York 11794--3800, USA}

\begin{abstract}
We discuss the generalized parton distributions (GPDs) for unpolarized hadrons, as a continuation of our recent
work on hadronic structure on the light front. We analyze the unpolarized GPDs for the light nucleon and delta, as well
as generic mesons, using the lowest Fock states. We use these GPDs to reconstruct the charge and gravitational 
form factors, and discuss their relative sizes. The results are also compared to reported QCD lattice results.
 \end{abstract}
\maketitle

\section{Introduction}
Light cone distributions are central to the description of hard inclusive and exclusive processes. Thanks to factorization, a hard process factors  into a perturbatively calculable contribution times pertinent parton distribution and fragmentation functions. Standard examples can be found in deep inelastic scattering, Drell-Yan process and jet production to cite a few.
	
The parton distribution functions (PDF) are forward matrix elements of the leading twist operators of pertinent light front wave functions. They are valued time-like and as such not readily amenable to lattice simulations. As a result, only few moments of the PDF have been accessible to numerical simulations. To overcome this difficulty, space-like quasi-parton distrribution functions (qPDF)  for fixed hadron momentum have been suggested by Ji~\cite{Ji:2013dva}, which are perturbatively matched to the light front PDF in  the large 	momentum limit.

The generalized parton distributions (GPD) are off-forward matrix elements of the leading quark and gluon operators on the light front.
They capture more aspects of the partonic content of the light front wavefunctions. They provide a more comprehensive description of the partons in a hadron on the light front, that range from its longitudinal momentum distribution (PDF), to its spatial charge and current distributions as captured by form factors (FF). GPDs are  accessible by semi-inclusive processes through deeply virtual Compton scattering (DVCS), and deeply virtual meson production~\cite{Ji:1996nm,Radyushkin:1997ki}. DVCS has been pursued by the CLAS collaboration at JLab and the COMPASS collaboration at CERN, with more planned experiments at the future electron ion colliders (EIC, EIcC)~\cite{AbdulKhalek:2021gbh,Anderle:2021wcy}.

Throughout, we will only discuss the GPDs in the DGLAP or large-x region $\xi\leq x\leq 1$, with $x$ the parton fraction of the struck quark,
and $\xi$ its longitudinal skewness, i.e. the fraction of light cone momentum transfer to the nucleon, with the total 4-momentum
transfer $\Delta$ and $\Delta^2=t<0$~\cite{Belitsky:2005qn} (and references therein).
This regime corresponds to the quark bag diagram in DIS kinematics. The GPDs in the ERBL 
or low-x region $0\leq x\leq \xi$ are not accessible with our LF wavefunctions. This regime corresponds to a particle changing diagram,
with different in-out Fock states. Because of this kinematical limitation, the important concept of Polynomiality cannot be checked
from our results. In the forward limit with $\xi=0$ and $\Delta=0$, the GDPs reduce to pertinent  PDFs, and integrate along
$x$ to form factors irrespective of $\xi$, bringing together the concepts of  parton densities and form factors.

The outline of this paper is as follows. In section~\ref{SEC_MESON} we recall the leading twist-2 definitions of the unpolarized
and polarized GPDs for generic mesons, and summarize the pertinent  kinematics. We also make explicit the unpolarized GPDs for the light 
and heavy mesons, using the generic form of the LF wavefunctions  established in our earlier work~\cite{}.
construct the leading and unpolarized GPD. In esction~\ref{SEC_BARYON} we also detail the unpolarized and polarized GPDs for 
generic and non-exotic baryons, with their relevant kinematics. We  carry the numerical analysis of the unpolarized GPD for the 
nucleon and $\Delta$-isobar, using the LF wavefunctions in~\cite{Shuryak:2021hng,Shuryak:2021mlh}.  We show that the nucleon shape at various parton-x scans
on the light front, is substantially different from the rest frame, especially for large-x. In section~\ref{SEC_FF} we show that the
electromagnetic Dirac form factor of both the nucleon and $\Delta$-isobar, are recovered from the GPD for zero skewness.  We
make explicit the A-gravitational formfactor, with a comparison to the Dirac formfactor. 
The results are also compared to available QCD lattice results.
Our conclusions are in section~\ref{SEC_CON}.

\section{Twist-2 GPDs of mesons}
\label{SEC_MESON}

The GPDs provide a complete description of the leading twist-2 quark and gluon substructure in QCD of a hadron.
They interpolate between the partonic densities and hadronic form factors, and as such provide a richer access to the
hadronic structure~\cite{Belitsky:2005qn} (and references therein).
More specifically, the  GPDs are the twist-2 spin-j matrix elements of local quark- and gluon-operators,
measured using LFWFs. At the resolution of about $\mu_0=1$ GeV, the LFWFs are composed of constituent quarks, with
the non-perturbative gluons giving rise to the constituent masses, string tension and non-perturbative  spin forces.

With this in mind, the leading and {unpolarized} GPD as a matrix element of solely  the quark  twist-2 spin-2 local operator, in
a generic meson state on the LF, is~\cite{Belitsky:2005qn}
\bea\label{GPD}
&&{H}(x,\xi, t)=
 \int \frac{dz^-}{4\pi}e^{i\frac{x}{2}P^+ z^-}\nonumber\\
&&\times \langle M(p_+)|\bar q (0)\gamma^{+} [0,z^-]q (z^-)|M(p_-)\rangle
\eea
with $z^-$ a time-like separation, for fixed $z^+=z_\perp=0$. The leading and {\it polarized} twist-2 meson GPD is
defined as~\cite{Belitsky:2005qn}
\bea\label{GPDE}
&&\frac {iq_\perp^i\epsilon_\perp^{ij}}{2m_M}{E}(x,\xi, t)=
 \int \frac{dz^-}{4\pi}e^{i\frac{x}{2}P^+z^-}\nonumber\\
 &&\times \langle M(p_+)|\bar q (0)i\sigma^{j+}\gamma^5  [0,z^-]q (z^-)|M (p_-)\rangle
\eea
 with $\epsilon_\perp^{ij}$ the antisymmetric tensor in the transverse plane with $i,j=1,2$, and $\sigma^{j+}=\frac i2[\gamma^j, \gamma^+]$.  $H$ is chirally even and $E$ is chirally odd. The latter probes the spatial distribution of a transversely polarized quark in the  boosted meson state. The GPD kinematics in the symmetric frame, is fixed  as follows
 \bea
&& p_\pm =P\pm \Delta/2, \qquad P=(P^0,0_\perp,P^z),\nonumber\\
&&P\cdot \Delta=0, \qquad  t=-\Delta^2, \qquad
P^2= -m_M^2+\frac 14 t, \nonumber\\
&& x=\frac{k^+}{P^+}, \qquad \xi=\frac{p_+^+-p_-^+}{p^+_++p_-^+}=-\frac{\Delta^+}{2 P^+}\,.
\eea
Note that we use the mostly-plus metric convention, so on-shell squared momenta are negative, and $t<0$. The light-front  longitudinal skewness is referred to 
as $\xi$. The Wilson link will be set as  $[0, z^-]\rightarrow 1$ throughout. 

\subsection{Mesons}
Any meson is characterized by three leading Fock state wavefunctions, that mix under $L_z, S_z$ on the LF. The
classification of the states  is done using $\Lambda=(S_1+S_2)_z+L_z$, so the labelling $\Lambda=0, \pm 1$. Although $\Lambda=\pm 1$ 
are tied by symmetry modulo a trivial azimuthal phase, we will keep the three-label assignment. For the net $\Lambda=0$ meson pseudoscalar ($P$) and vector ($V$)
states we have~\cite{Ji:2003yj}
 \begin{widetext}
\bea
 \label{WFPS}
	| P \rangle& =&\int d[1]d[2]{\delta_{ij} \over \sqrt{N_c}} \ \big[ 
	\psi_0^{P}(x,k_\perp) 
	\big( Q_{i\uparrow}^\dagger (1) \bar Q_{j\downarrow}^\dagger(2)- 
	Q_{i\downarrow}^\dagger(1) \bar Q_{j\uparrow}^\dagger (2)\big) \nonumber \\
&&\qquad\qquad\qquad +
	 \psi_{-1}^{P}(x,\vec k_\perp) 
	Q_{i\uparrow}^\dagger  (1) \bar Q_{j\uparrow}^\dagger(2)+ 
		\psi_{+1}^{P}(x,\vec k_\perp)  Q_{i\downarrow}^\dagger(1) \bar Q_{j\downarrow}^\dagger (2)
	\big] |0\rangle
	\eea
	\bea
 \label{WFVX}
	| V \rangle& =&\int d[1]d[2]{\delta_{ij} \over \sqrt{N_c}} \ \big[ 
	\psi_0^V(x,k_\perp) 
	\big( Q_{i\uparrow}^\dagger (1) \bar Q_{j\downarrow}^\dagger(2) +
	Q_{i\downarrow}^\dagger(1) \bar Q_{j\uparrow}^\dagger (2)\big) \nonumber \\
&&\qquad\qquad\qquad +
	\psi_{-1}^V(x,\vec k_\perp)Q_{i\uparrow}^\dagger  (1) \bar Q_{j\uparrow}^\dagger(2) -
		\psi_{+1}^V(x,\vec k_\perp) Q_{i\downarrow}^\dagger(1) \bar Q_{j\downarrow}^\dagger (2)
	\big] |0\rangle
\eea
\end{widetext}
with $N_c=3$, for the pseudoscalar and vector respectively. The subscripts $0$ and $\pm 1$ on the wave-functions, refer to  $L_z$, the z-projections of the orbital momentum. Note that compared to the notations in \cite{Ji:2003yj},
there are no explicit factors of $k_\perp^{\pm}=k_1\pm ik_2$ here because
they naturally belong to our wave functions,  consistently defined not only for $m=L_z=1$,  but for any $m$ value. Inserting (\ref{WFPS}) into (\ref{GPD}-\ref{GPDE})
and carrying the contractions  yields for the pseudoscalar P-state and unpolarized GPD, at zero skewness
\bea
\label{HPSEUUNPO}
H(x,0,t)=\sum_{\Lambda=0, \pm 1} \int \frac{dk_\perp}{(2\pi)^3}\psi_\Lambda^{P*}(x, k_\perp^\prime)\psi_\Lambda^P(x, k_\perp)\nonumber\\
\eea
while for the polarized P-state and zero skewness ($j=1,2$)
\begin{widetext}
\bea
\label{HPSEUPOL}
&&\frac{i\Delta^j_\perp}{2m_M} E(x,0,t)=\int \frac{dk_\perp}{(2\pi)^3}\nonumber\\
&&\bigg((-i)^j\bigg(\psi^{P*}_{-1}(x, k_\perp^\prime)\psi^P_{0}(x, k_\perp) + \psi^{P*}_{0}(x, k^\prime_\perp) \psi^P_{+1}(x, k_\perp) \bigg)
+(+i)^j\bigg(\psi^{P*}_{+1}(x, k^\prime_\perp) \psi^P_{0}(x, k_\perp) +\psi^{P*}_{0}(x, k_\perp^\prime)\psi^P_{-1}(x, k_\perp) \bigg)\bigg)\nonumber\\
\eea
The kinematical arrangement is as follows: a/
{\bf active quark}:  $k_\perp^\prime=k_\perp +(1-x)\Delta_\perp$;
b/ {\bf passive quark}:  $k_\perp^\prime=k_\perp - x \Delta_\perp$. The transferred momentum to the meson state is $t=-\Delta_\perp^2$.

More specifically, we will assume that the struck quark is the one with
momentum fraction $x$, and that the kick momentum $\vec \Delta_\perp$ is in transverse 1-direction.
 Then  after the kick,  the  the longitudinal and transverse momenta in the 1-direction (in primed notations) are 
 \bea
&& x_1^\prime={x-\xi \over 1-\xi},\,\,\, x_2^\prime=1-x\nonumber\\
&& k_1^\prime=k+\Delta_\perp {1-x \over 1-\xi },\,\, k_2^\prime=k- \Delta_\perp {x  \over 1-\xi }
 \eea
 Our detailed analyses in~\cite{Shuryak:2021hng,Shuryak:2021mlh} show that the mesonic LF wavefunctions
can be well approximated by a Gaussian 
$\psi(k,x)\sim e^{-A(x)(k_1^2+k_2^2)}$, where $A(x)$ depends on the specific nature of the meson. As a result, 
the meson LF wavefunction after the kick $\psi(k^\prime,x^\prime)$ follows through the substitution
 $(k_1^2+k_2^2) \rightarrow (k_1'^2+k_2'^2)$. Using (\ref{HPSEUUNPO})
to calculate the unpolarized GPD $H(x,0,-\Delta_\perp^2)$ at zero skewness, yields
\be 
H(x,0,-\Delta_\perp^2) \sim e^{- A(x)\big(3/4-x(1-x)\big)\Delta_\perp^2 }
\ee

\section{Twist-2 spin-2 GPDs of the nucleon}	
\label{SEC_BARYON}
The leading nucleon GPDs are also driven by the leading twist-2 and spin-2 vector and axial-vector currents on the light front.
As we noted earlier for the mesons, our construction of the nucleon LFWFs restricts our analysis of the GPDs to 
the DGLAP region, with particle preserving in-out Fock states. Also at the resolution of $\mu_0=1$ GeV, the nucleon is
limited to the lowest 3-quark Fock state, where the GPDs are limited to their constituent quark content. 
In this section, we will quote the general results for
the unpolarized and polarised GPDs for generic baryons. The unpolarized GPD for the nucleon and $\Delta$-isobar 
will be made more explicit using our LFWFs in~\cite{Shuryak:2022thi,Shuryak:2022wtk}.

\subsection{General expressions and kinematics}
The  quark GPDs of the nucleon are captured by the off-diagonal formfactors~\cite{Belitsky:2005qn}
\bea
\label{GPD1}
&&\int\frac{P^+ dz^-}{4\pi}e^{\frac 12 ixz^-P^+}
\langle p_+\Lambda_+|\overline q(0)\gamma^+[0,z^-] q(z^-)|p_-\Lambda_-\rangle=\nonumber\\
&&\overline{N}(p_+,\Lambda_+)\bigg(H(x, \xi, t)\gamma^++E(x,\xi, t)\frac{i\sigma^{+j}\Delta_j}{2m_N}\bigg) N(p_-,\Lambda_-)\nonumber\\
&&\int\frac{P^+ dz^-}{4\pi}e^{\frac 12 ixz^-P^+}
\langle p_+\Lambda_+|\overline q(0)\gamma^+\gamma_5[0,z^-] q(z^-)|p_-\Lambda_-\rangle=\nonumber\\
&&\overline{N}(p_+,\Lambda_+)\bigg(\tilde{H}(x, \xi, t)\gamma^+\gamma_5+\tilde{E}(x,\xi, t)\frac{\Delta^+\gamma_5}{2m_N}\bigg) N(p_-,\Lambda_-)
\eea
Here $H,E$ are the unpolarized quark GPDs, and $\tilde H, \tilde E$ their polarized counterparts. Note that for $\xi=0$, we have $t=-\Delta_\perp^2$,
and $\tilde E$ drops out. The generic form of the nucleon wavefunction is
\bea
\label{GPD2}
|p\Lambda\rangle= \int \prod_{i=1}^3\frac{dx_i dk_{i\perp}}{\sqrt{x_i}}
\delta\bigg(1-\sum_{i=1}^3x_i\bigg)\, \delta\bigg(\sum_{i=1}^3k_{i\perp}\bigg)
\psi^{\Lambda} ([x_i, k_{i\perp},\lambda_i])|[x_ip^+, k_{i\lambda}+x_ip_{i\perp}, \lambda_i]\rangle
\eea
for a nucleon of total momentum $p^\mu$ and helicity $\Lambda=\pm 1$.	The conversion to the Jacobi coordinates is subsumed, with the delta-functions
readily enforced. At low resolution, the nucleon state $\psi^\Lambda$  is a quark-diquark $q[qq]_0$ Fock state.

To proceed, we note the identities for the matrix elements of the nucleon spinors in (\ref{GPD1}) on the right-hand-side,
\bea
\label{NSPINORS}
\overline N(p_+,\Lambda_+)\gamma^+N(p_-\Lambda_-)&=&2P^+\frac{\sqrt{1-\xi}}{1-\frac 1 2\xi}\,\delta_{\Lambda_+\Lambda_-}\nonumber\\
\overline N(p_+,\Lambda_+)\frac{i\sigma^{+j}\Delta_j}{2m_N}N(p_-\Lambda_-)&=&2P^+
\bigg(-\frac 14 \frac{\xi^2}{(1-\frac 12 \xi)\sqrt{1-\xi}}\delta_{\Lambda_+\Lambda_-}+
\frac{\Lambda_- q^1+iq^2}{2m_N}\frac 1{\sqrt{1-\xi}}\delta_{(-\Lambda_+)\Lambda_-}\bigg)
\eea
Inserting the free field decomposition of the
good component of the quark field in (\ref{GPD1}), and carrying the contractions yield the respective GPDs.
\\
\\
{\bf Zero skewness:}
\\
\bea
H(x,0,t)&=&\int_P \delta(x-x_1)\,\psi^{+*} ([x^\prime_i, k^\prime_{i\perp},\lambda_i])\psi^{+} ([x_i, k_{i\perp},\lambda_i])\nonumber\\
E(x,0,t)&=&\frac{-2m_N}{q_L}\int_P \delta(x-x_1)\,\psi^{+*} ([x^\prime_i, k^\prime_{i\perp},\lambda_i])\psi^{-} ([x_i, k_{i\perp},\lambda_i])\nonumber\\
\tilde H(x,0,t)&=&\int_P\lambda_1\, \delta(x-x_1)\,\psi^{+*} ([x^\prime_i, k^\prime_{i\perp},\lambda_i])\psi^{+} ([x_i, k_{i\perp},\lambda_i])\nonumber\\
\eea
with $q_L=q^1-iq^2$, and the phase space helicity sum and integration
\bea
\int_P=\sum_{[\lambda_i]}\int \prod_{i=1}^3dx_id^2k_{i\perp}\,\delta\bigg(1-\sum_{i=1}^3x_i\bigg)\, \delta\bigg(\sum_{i=1}^3k_{i\perp}\bigg)
\eea
The kinematical arrangement is as follows: a/
{\bf active quark} $i=1$:  $x_1^\prime=x_1$ and $k_{1\perp}^\prime=k_{1\perp}+(1-x_1)\Delta_\perp$ and $\lambda_1=\pm 1$;
b/ {\bf passive quarks} $i=2,3$:  and $x_i^\prime=x_ i$ and $k_{i\perp}^\prime=k_{1\perp}-x_i\Delta_\perp$.
\\
\\
{\bf Finite skewness:}
\\
\bea \label{eqn_for_H}
\frac{\sqrt{1-\xi}}{1-\frac 12 \xi} H(x, \xi, t)- \frac 14 \frac{\xi^2}{(1-\frac 12 \xi)\sqrt{1-\xi}} E(x, \xi, t)=
\int_P \delta(x-x_1)\,\psi^{+*} ([x^\prime_i, k^\prime_{i\perp},\lambda_i])\psi^{+} ([x_i, k_{i\perp},\lambda_i])
\eea
and
\bea
-\frac {q_L}{2m_N}\frac 1{\sqrt{1-\xi}} E(x, \xi, t)=
\int_P \delta(x-x_1)\,\psi^{+*} ([x^\prime_i, k^\prime_{i\perp},\lambda_i])\psi^{-} ([x_i, k_{i\perp},\lambda_i])
\eea
The kinematical arrangement is as follows: a/
{\bf active quark} $i=1$:  $x_1^\prime=(x_1-\xi)/(1-\xi)$ and $k_{1\perp}^\prime=k_{1\perp}+((1-x_1)/(1-\xi))\Delta_\perp$ and $\lambda_1=\pm 1$;
b/ {\bf passive quarks} $i=2,3$:  and $x_i^\prime=x_ i$ and $k_{i\perp}^\prime=k_{1\perp}-(x_i/(1-\xi))\Delta_\perp$.

\end{widetext}

\subsection{$N, \Delta$ GPDs from the LFWFs with $\xi=0$}
\label{SEC_FF}
The unpolarized GPDs for the nucleon and $\Delta$-isobar can be explicitly
evaluated using the the explicit LFWFs developed in~\cite{Shuryak:2022wtk},
to which we refer for most of the details. Here, we briefly recall for the flavor
and spin symmetric isobar $\Delta^{++}=uuu$, the LF Hamiltonian is the sum of the
kinetic energies of three u-constituents, plus their confining potentials (the hyperfine
spin forces are smal and can be added in perturbation). The proton $uud$ is composed 
of a flavor-spin asymmetric $[ud]$ diquark  paired by the instanton-induced $^\prime$t Hooft 
interaction, treated in the quasi-local approximation. This diquark pairing is responsible for the 
mass splitting between the nucleon and the $\Delta$-isobar, and the difference between thei
respective LFWFs.

For the proton with quark assignment $uud$, we assume that the struck quark is $d$, with longitudinal 
momentum fraction $x_3$. In our (modified) Jacobi coordinates, this momentum fraction is directly related to the
longitudinal variable $\lambda$.  For the unpolarized d-quark GPD in (\ref{eqn_for_H}), this amounts to integrating
the off-forward LFWFs over 5 variables,  the transverse momenta $\vec p_\rho,\vec p_\lambda$ and  
$\rho$  for the nucleon. To proceed, we approximate the dependence on the transverse momenta 
$\vec p_\rho,\vec p_\lambda$ by Gaussians, which is quite accurate, and carry the integrals analytically. 
The remaining integration over $\rho$ is performed numerically. We recall that the LFWFs are generated
at a  low renormalization scale, say $\mu_0=1$ GeV, with a nucleon composed of three constituent quarks, without
 constituent gluons. (The nonperturbative vacuum gluonic fields and hard gluons are all repackaged in the constituent quark mass and chiral condensate
on the LF).

\begin{figure}[htbp]
\begin{center}
\includegraphics[width=8cm]{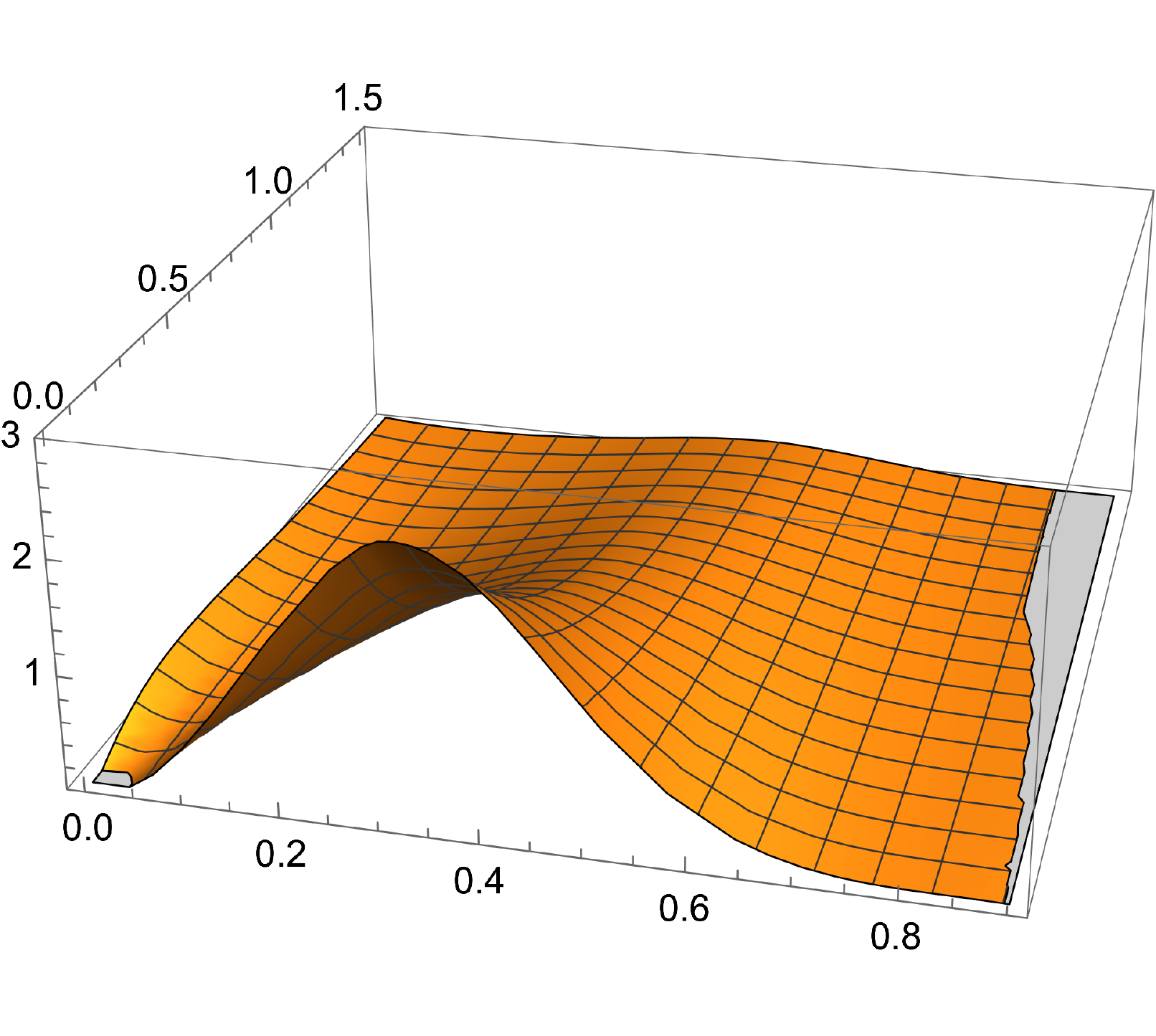}
\caption{The nucleon GPD function $H_d^N(x,\xi=0,Q^2)$ for a struck d-quark.}
\label{fig_Hd_N_3d}
\end{center}
\end{figure}

In Fig.\ref{fig_Hd_N_3d} we show the unpolarized nucleon GPD for the struck d-quark,
 as a function of $x,Q^2$ and zero skewness.  At small $Q^2$, the dependence
 on the longitunal parton momentum $x$, is that expected for a PDF with a maximum at 
 $x=\frac 13$. At  larger $Q^2$, the maximum of the  GPD clearly shifts towards
larger values of $x$. The GDP is not simply factorizable into the PDF times the form
factor, which are separable in $x$ and $Q^2$. The right shift in Fig.\ref{fig_Hd_N_3d}
shows that the nucleon shape changes with $x$. This is a key point of interest to us,
as we now proceed to detail these shape modifications.

Theoretical considerations~\cite{Guidal:2004nd} have suggested that the GPDs can be approximated
generically, by a  ``Gaussian ansatz"  in the momentum transfer $Q^2$, with a width and a
pre-exponent that are x-dependent
\bea
 {\rm GPDs}(x,Q^2) \sim f_1(x)\,e^{-Q^2 f_2(x)}
 \label{GPD_ansatz}
\eea
We recall that  the standard nucleon formfactors are  dipole-like at low $Q^2$
(say lower than 10 GeV$^2$) with $F(Q^2)  \sim 1/(m^2+Q^2)^2$. At  very large $Q^2$
it asymptotes to a constant  $Q^4 F(Q^2) \rightarrow const$  which is fixed by the 
perturbative QCD scattering rule. In between, the $Q^2$ dependence remains an open
issue. 

Our LF formfactors were found to be consistent with these observations
(see below), and our GPDs at fixed $x$  are indeed numerically consistent with the exponent
\bea
\label{APPROX}
e^{-\frac{Q^2 f_s(x)}{\sqrt{1-\xi^2}}}
\eea
with no dependence of the pre-expont on the skewness, for $x>\xi$.
We have checked that the x-integration of this exponent with $\xi=0$, returns the expected rational formfactor.

\begin{figure}[htbp]
\begin{center}
\includegraphics[width=6cm]{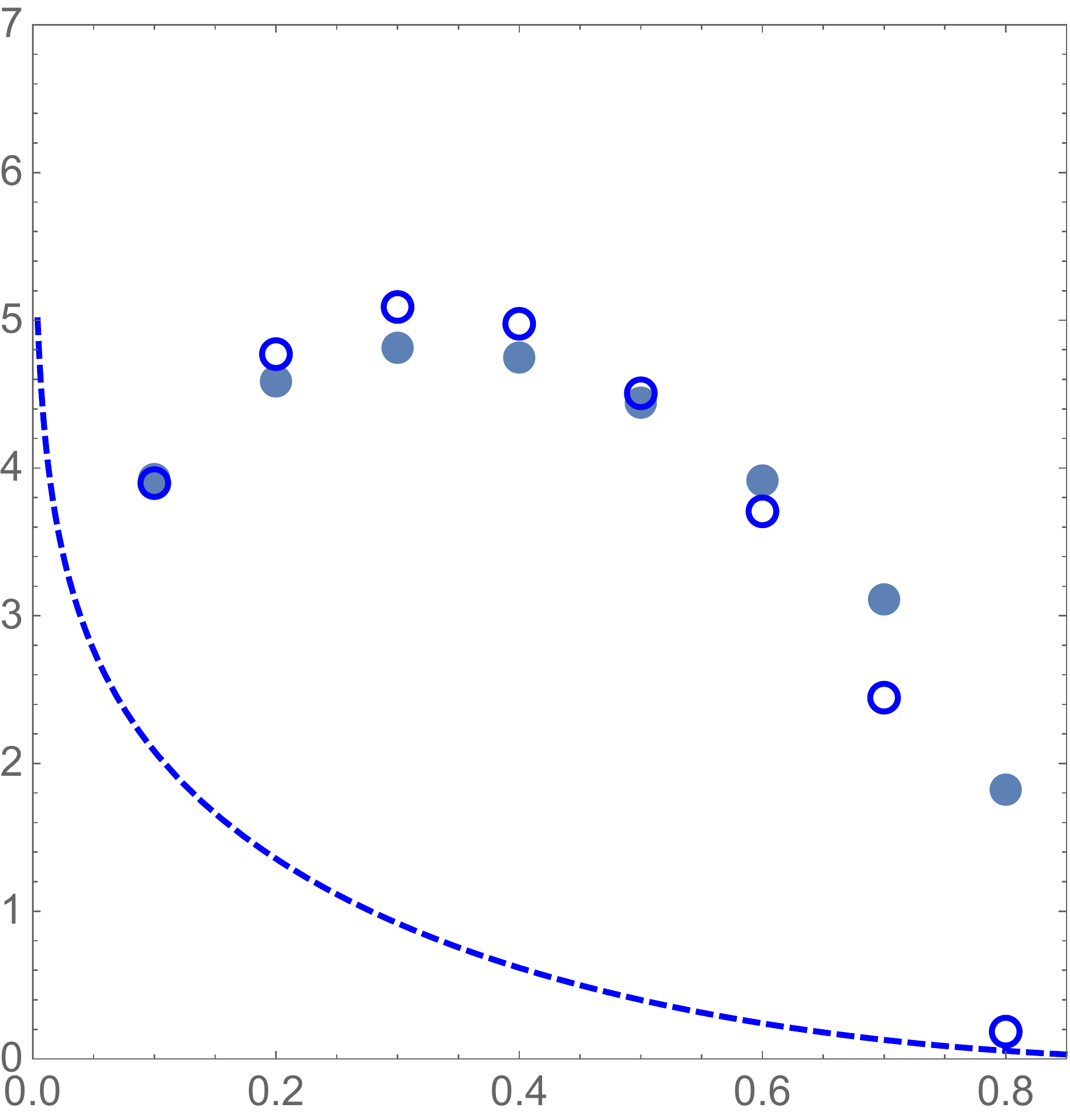}
\caption{Closed and open points  correspond to  slopes $-d log[{\rm GPD}]/dQ^2$ (at $Q^2\approx 1\,{\rm GeV}^2$) calculated from our LFWF for the nucleon and Delta baryons, respectively.
The dashed line shown for comparison is the ansatz (\ref{eqn_GPD_ansatz}) .
}
\label{fig_f_of_x_GPD}
\end{center}
\end{figure}

The GPDs calculated from our LFWFs  is well described by (\ref{GPD_ansatz}).
In particular, the r.m.s. spatial size of the struck d-quark  for fixed $x$ is
\bea
R_{r.m.s.}(x)=\sqrt{2 f_2(x)} 
\eea
In Fig.\ref{fig_f_of_x_GPD} we show the effective Q-slope of the unpolarized 
d-quark GPD,  for the nucleon (filled-points) and $\Delta$-isobar (open-points).
The r.m.s. size $R_{r.m.s.}(x)$ is maximal at $x\sim \frac 13$, and
is numerically about $0.6$ fm. It decreases sharply for $x\sim 1$, and
moderatly for $x\sim 0$. This can be explained by the fact that for $x\sim \frac 13$,
all three quarks carry about the same longitudinal momentum on the LH, which 
corresponds to three quarks at rest in the CM or rest frame. 
Semiclassically, this means a configuration in which a struck quark is near a ``turning points" of
the wave function, with their QCD strings maximally streched. In contrast, $x$
away from $\frac 13$, corresponds to a struck quark rapidly moving in the CM frame,
which must happen near the hadron center, The corresponding size is therefore small.
 Note that for $x\rightarrow 1$, the distribution of the struck d-quark becomes nearly pointlike,
with slopes close to zero.  Our LFWFs show that the magnitude of this effect is 
different for nucleon and $\Delta$-isobar,   sensitive to the the diquark substructure of the former.



The   behavior of the GPDs near the edges $x\sim 0,1$, can be gleaned
from general QCD considerations. The small x-region of the GPD is
dominated by Regge physics~\cite{Guidal:2004nd},   while the large x-region of the
GDP is fixed by the  Drell-Yan-West relation.  A particular functional form for the GPD 
that abides by these two limits, was suggested in~\cite{deTeramond:2018ecg} 
\ba 
\label{eqn_GPD_ansatz}
H(x,0,t)&\sim &e^{t f(x) }
\ea
with $t=-Q^2$, and 
$$f(x)={1 \over 4\lambda} \bigg((1 - x)\,{\rm ln}\bigg({1\over x}\bigg)+ a(1 - x)^2\bigg)$$
with the  parameters $a=.53$ and $ \lambda=(0.548\, {\rm GeV})^2$. 
In Fig.\ref{fig_f_of_x_GPD} we show (\ref{eqn_GPD_ansatz}) (dashed-line)
for comparison.  The sharp rise at small $x$ is due to the parametrized Reggeon
in (\ref{eqn_GPD_ansatz}),
 a multi-parton cloud around a baryon. It is absent in our approach, which is
 limited to the lowest Fock state. The decrease at large $x$ is in qualitative 
 agreement with our results. However, at intermediate $x$ there is a significant 
 disagreement, with our baryon  r.m.s. size $R_{r.m.s.}(x)$, which is significantly larger.

\section{Electromagnetic and gravitational  form factors}
\label{SEC_FF}
On the LF the GPDs  are related to various  form factors of the nucleon. In particular, the n-Mellin moments of the GDP is a polynomial 
of degree $\xi^n$, a property known as polynomiality~\cite{Belitsky:2005qn}. However, since the Mellin moments sum over both the ERBL region 
($0<x<\xi$) and the DGLAP region ($\xi<x<1$), the polynomiality cannot be checked, as the ERBL region falls outside the scope of our analysis. 

\subsection{Electromagnetic formfactors}
\begin{figure}[t]
\begin{center}
\includegraphics[width=8cm]{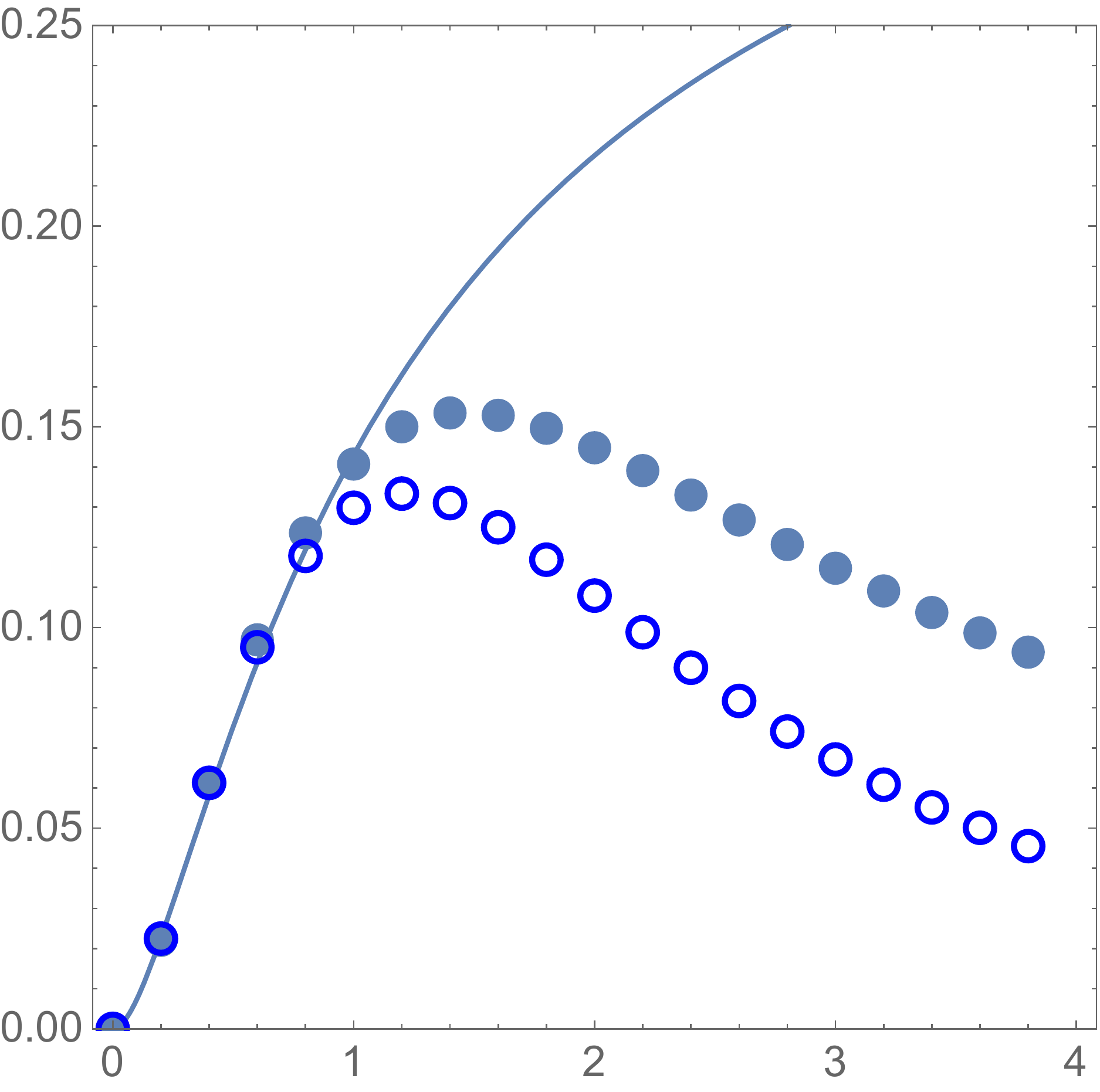}
\caption{Scaled Dirac form factor $Q^4 F_1^d(Q^2)$ in GeV$^4$ versus $Q^2$ in GeV$^2$, for the nucleon (filled-point) and $\Delta$ (open-point), obtained by integrating  the unpolarized d-quark-H GDP. For comparison we show a ``dipole fit" curve $Q^4/(1+Q^2/m_\rho^2)^2$ (solid-line).}
\label{fig_ff_N_del}
\end{center}
\end{figure}

This notwithstanding, a number of electromagnetic and gravitational formfactors of the nucleon, can be extracted from the present GPDs,
allowing also for estimates of the spin and mass sum rules. More specifically, the Dirac $F_1$, the Pauli $F_2$ electromagnetic formfactors 
and the axial formfactor $G_A$,  are all  tied to the zeroth moment of the GPDs on the LF, at zero skewness~\cite{Belitsky:2005qn} (and references therein)
\bea
F_1(t)&=&\int dx H(x,0,t)\nonumber\\
F_2(t)&=&\int dx E(x,0,t)\nonumber\\
G_A(t)&=&\int dx \tilde H(x,0,t)
\eea
In  Fig.\ref{fig_ff_N_del} we show the numerical results for $Q^4F^d_1$ versus $Q^2$ in ${\rm GeV}^2$,  
for a struck d-quark, following from the integration of the  
unpolarized d-quark $H$ GPD, for the nucleon (filled-point) and $\Delta$-isobar (open-point). The standard dipole-form factor with
the rho mass is depicted by the solid-curve, using $Q^4/(1+Q^2/m_\rho^2)^2$ and $m_\rho=780$ MeV. The results follow the dipole curve 
below 1 GeV$^2$, and deviate substantially above, to asymptote a constant at much larger $Q^2$, as expected from the QCD counting rules
for both nucleons. The rescaled formfactor for the isobar is found to fall faster than the nucleon at large $Q^2$, which indicates that the nucleon
is more compact electromagnetically than the isobar, with a  smaller  electromagnetic radius.

\begin{figure}[h!]
\begin{center}
\includegraphics[width=5.5cm]{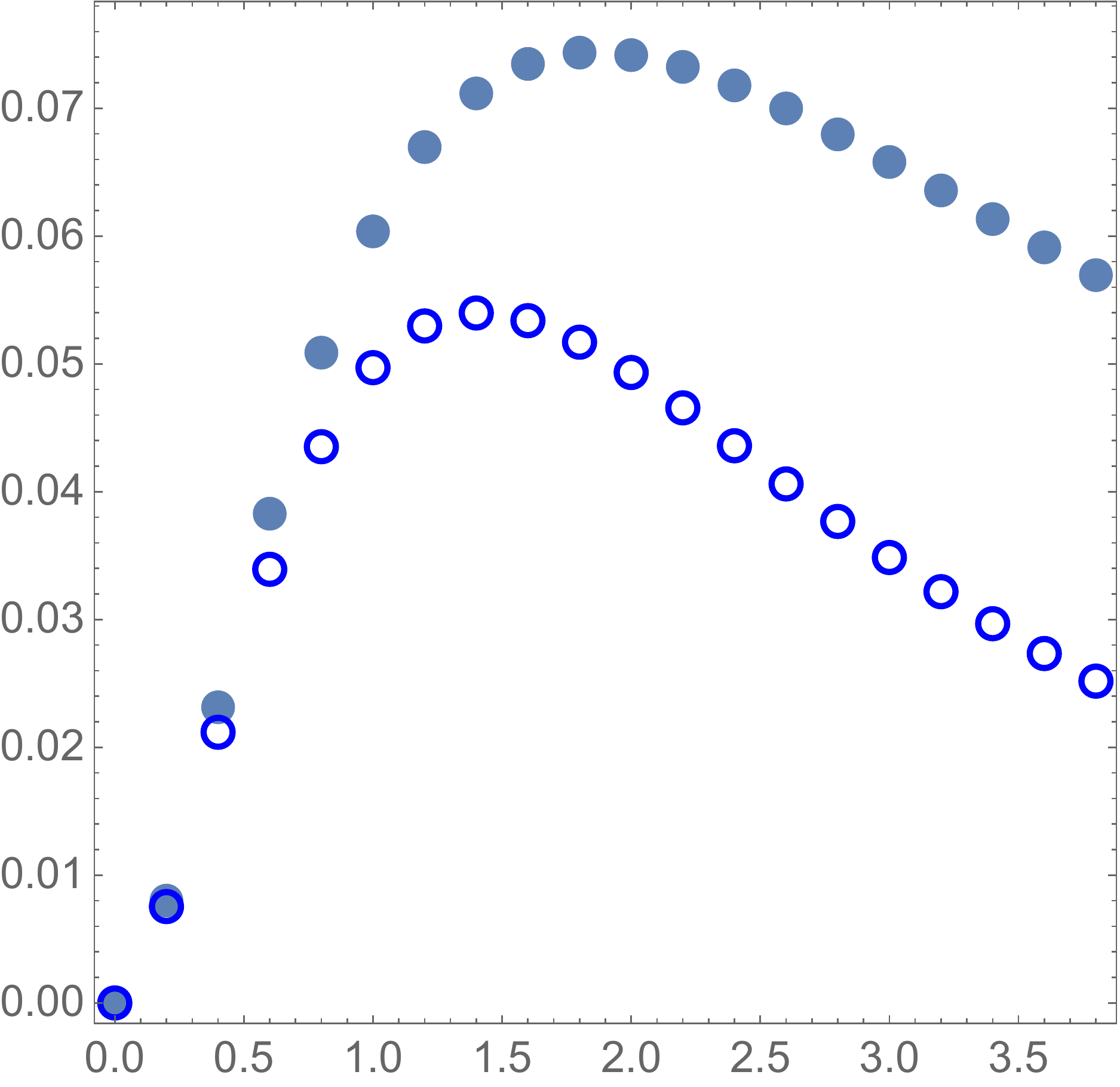} 
\includegraphics[width=5.5cm]{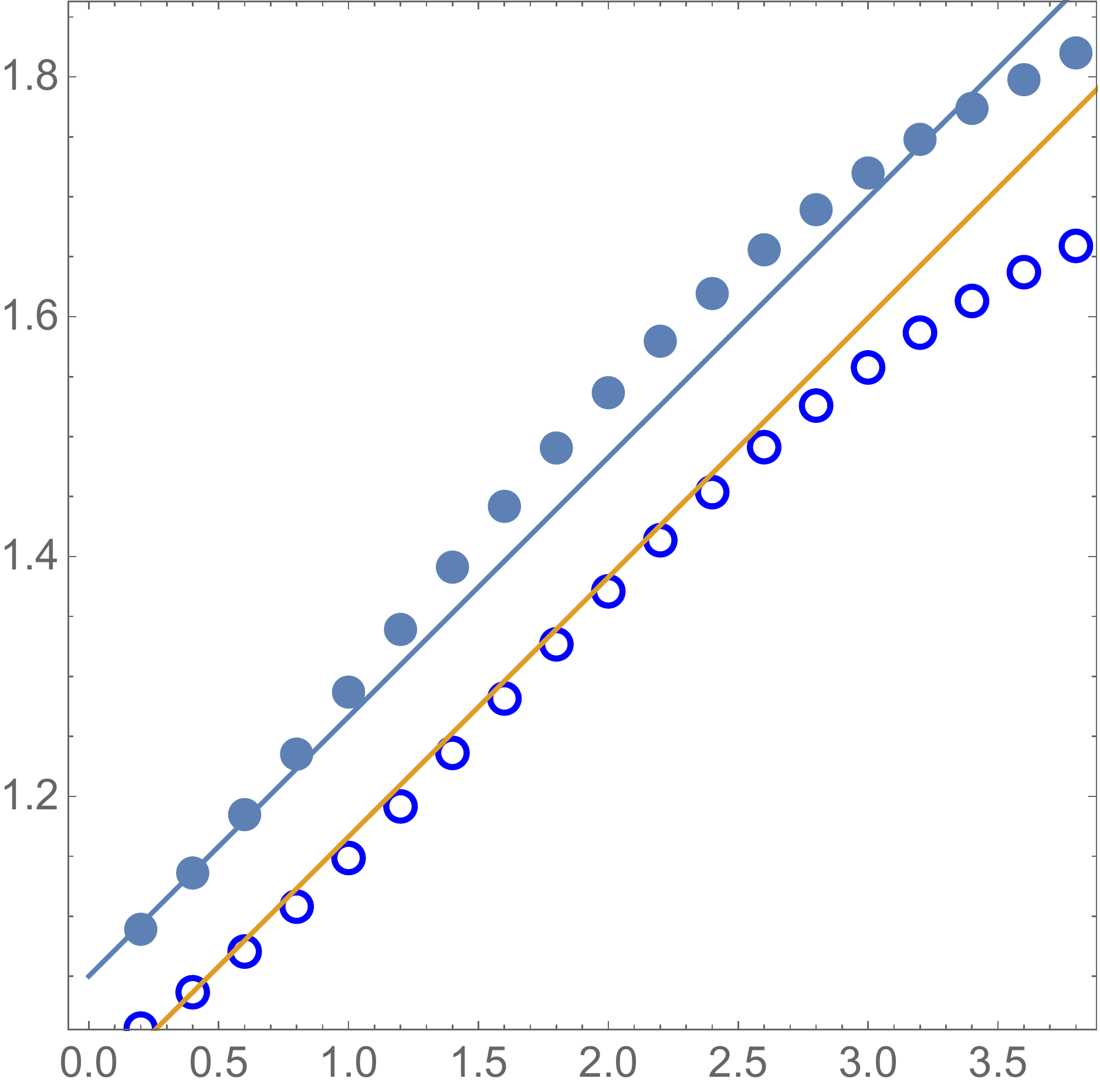} 
\caption{Scaled gravitational form factor $Q^4 A_d(Q^2)$ in GeV$^4$ versus $Q^2$ in GeV$^2$ (top), for the struck d-quark in a  nucleon (closed-points) and the isobar (open-points). Ratio of the gravitational to electromagnetic formfactors $3 A^d(Q^2)/F_1^d(Q^2)$ versus $Q^2$ in GeV$^2$ (bottom),  for the nucleon (closed-points) and the isobar (open points).
}
\label{fig_A_N_del}
\end{center}
\end{figure}

\subsection{Gravitational formfactors}
 As we noted earlier, our low-resolution LFWFs are dominated by the lowest
Fock state of three constituent quarks. Hence, the GPD is mostly that of the constituent quarks.
The first moment of the unpolarized GPDs at zero skewness,
is tied to the quark $A,B$ form factors of the energy-momentum tensor~\cite{Belitsky:2005qn} 
\bea
A(t)&=&\int dx \,x H(x,0,t)\nonumber\\
B(t)&=&\int dx \, xE(x,0,t)
\eea
They can be used to quantify the distribution of momentum, angular momentum and pressure-like stress,
inside the nucleon~\cite{Polyakov:2018zvc} (and references therein). More specifically, the total nucleon angular momentum at this low-resolution, is
given by Ji$^\prime$s sum rule~\cite{Ji:1996ek}
\bea
J=\frac 12=A(0)+B(0)
\eea
with the non-perturbative gluons implicit in the balance, as they enter implicitly  in the composition of the LF
Hamiltonian (mass, string tension, ...) for the constituent quarks. Since our LFWFs are so far unpolarized, we do 
not have acces to the B-formfactor, as  it involves the overlap between spin flipped LFWFs. The polarized GPDs together with the role of the spin forces, will be
discussed elsewhere.

In  Fig.\ref{fig_A_N_del} (top) we show the numerical results for $Q^4A^d$ versus $Q^2$ in ${\rm GeV}^2$,  
for a struck d-quark, following from the integration of the  
unpolarized d-quark-A GPD, for the nucleon (filled-point) and $\Delta$-isobar (open-point). Again, we observe
that the isobar form factor falls faster than the mucleon form factor, an indication that the nucleon is more compact gravitationally
than the isobar, with a smaller gravitational radius. In Fig.\ref{fig_A_N_del} (top) we plot the ratio of the gravitational formfactor 
relative to the electromagnetic form factor, for the struck d-quark in the nucleon (filled-points) and isobar (open-points). The decrease in $Q^2$ of the gravitational form factor, is slower than the electromagnetic form factor  for both hadrons. This means that the spatial mass distribution of the struck d-quark, is more compact  than the spatial charge  distribution.
More specifically, we find 
\bea
 \label{eqn_linear_in_ratio}
{3  A_d(Q^2) \over F^d_1(Q^2)}= C_0+{Q^2\over M^2_{fit}},
\eea
with slopes close to one $C_0^N=1.05$, $C_0^\Delta=0.95$, $M_{fit}^N=M_{fit}^\Delta=2.15$ GeV.  We have multiplied
$A^d$ by  a factor of 3, which accounts for the mean momentum $\langle x \rangle=1/3$, to bring the 
ratio close to 1.

\subsection{Comparison to lattice simulations}
The moments of the GPDs at  different momentum transfer $t$,  were evaluated on the lattice. Here, 
we will follow the detailed analysis by the LHPC collaboration~\cite{LHPC:2007blg}, 
where detailed numerical tables for the moments of the GPDs are given. More specifically, the 
longitudinal moments of the unpolarized 
nucleon GPD are defined as
\be H^n(\xi,t)=\int dx x^{n-1}H(x,\xi,t)  \ee
In their notations, the Dirac and gravitational formfactors at zero skewness $\xi=0$, are 
denoted by $A_1(0,t)$ and $A_2(0,t)$ respectively.
 Before we start comparing their results to ours, 
several warnings are in order.\\
(1) Although the simulations are done with domain wall fermions, they are still
done with large quark masses. The pion mass varies between $m_\pi=760$ and $350\, {\rm MeV}$ for different datasets. 
The chiral extrapolation to a small physical mass is clearly nonlinear, and produce
significant uncertainties.\\
(2) All reported results are quoted at a normalization scale $\mu^2=4\, {\rm GeV}^2$,
a standard value used for internal and external comparison. As we repeatedly emphasized
in the previous papers of this set, our LF wave functions and GPDs should correpond
to a much lower normalization point, and even with ``chiral evolution" are only taken up to
$\mu^2_{\chi}=1\, {\rm GeV}^2$. The difference between them is significant: if at $\mu_{\chi}$
there are no gluons and the quark fraction of momentum is 1, at $\mu^2=4\, {\rm GeV}^2$
it is about twice smaller, and comparable to the gluon momentum fraction. As the famous
``spin crisis" shows, a similar observation holds for the spin fractions carried by the quarks and gluons.\\
(3) The lattice spacing $a$ strongly limits the value of the largest momentum transfer which can
be used, to roughly $|t|<1.2 \, {\rm GeV}^2$. As one can see from our previous plots, interesting
deviations from simple dipole fits only are visible at larger values.\\ 

\begin{figure}[h]
\begin{center}
\includegraphics[width=6cm]{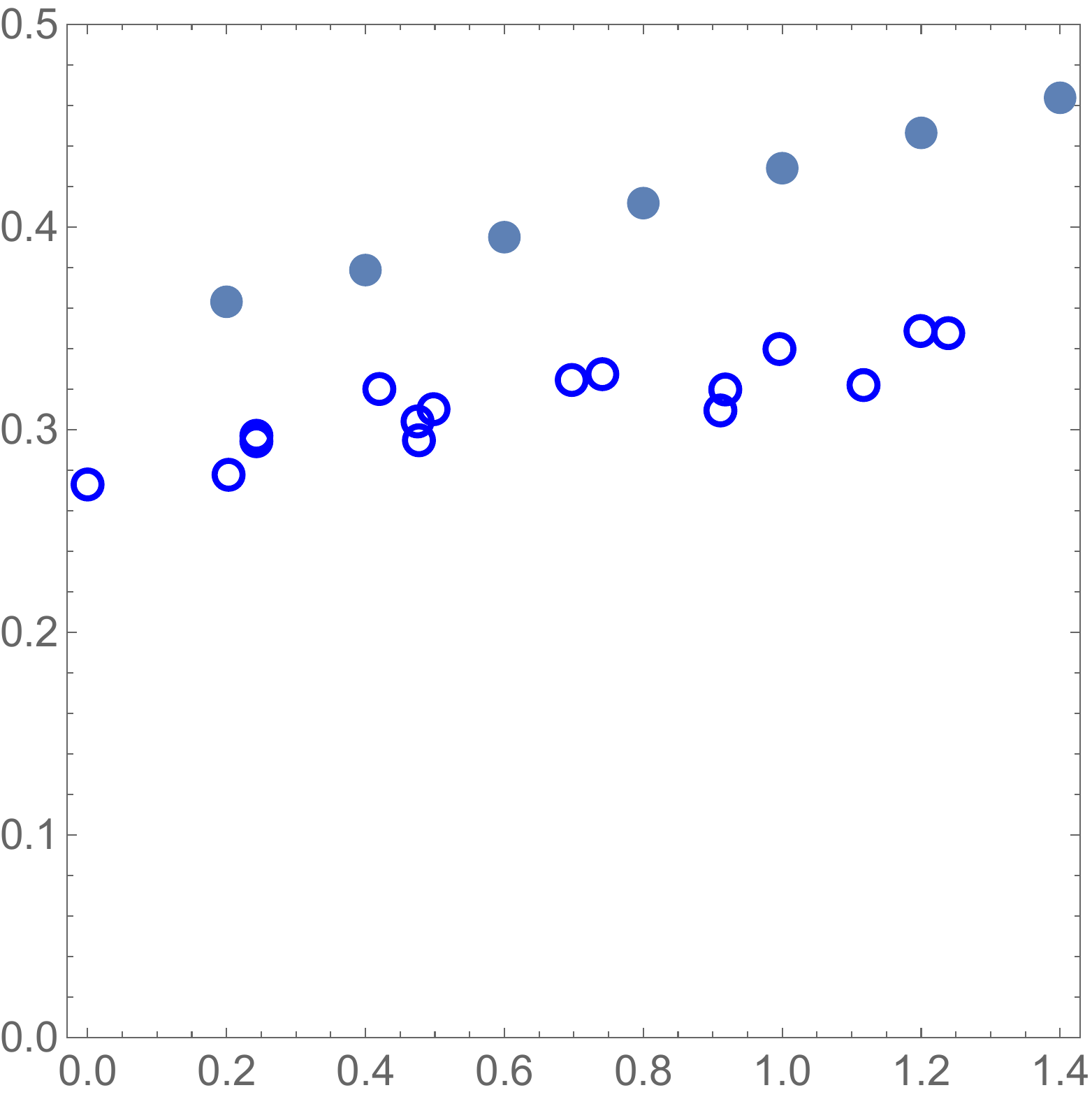}
\caption{ The  ratio $A_2(0,|t|)/A_1(0,|t|)$   as a function
of $|t|\, (GeV^2)$ for our  isoscalar $u+d$  nucleon GPD (filled points), and the lattice dataset 1 (open points) from \cite{LHPC:2007blg}.
}
\label{fig_FF_ratio_with_lattice}
\end{center}
\end{figure}

With these issues in mind, we now address the main qualitative findings. Perhaps the
most important observation is that the Dirac formfactor $A_1(0,t)$ decreases faster with  $|t|$,  than the gravitational formfactor $A_2(0,t)$.
This implies that spatial distribution of quarks in the nucleon is wider than that of the stress tensor. While this qualitative phenomenon is by now 
established empirically~\cite{Duran:2022xag}, the lattice data~\cite{LHPC:2007blg} quantify
it. In Fig.\ref{fig_FF_ratio_with_lattice} we compare  $t$-dependence for the ratio $A_2(0,|t|)/A_1(0,|t|)$, 
for our isoscalar $u+d$ nucleon GPD  (filled points), with the lattice dataset 1 from~\cite{LHPC:2007blg} (open points).

The main take from this comparison is that the ratio grows with the momentum transfer,
and the growth slope is similar for both results. However, the values of the ratio itself are not the same.
This is  expected, since  the evolution of our results (filled points)  to higher  chiral scale $\mu^2_{\chi}=1\, {\rm GeV}^2$ should cause the quark 
PDFs to shift to smaller $x$, therefore lowering down our curve  towards the reported lattice data points (open points).

\begin{figure}[h]
	\begin{center}
		\includegraphics[width=6cm]{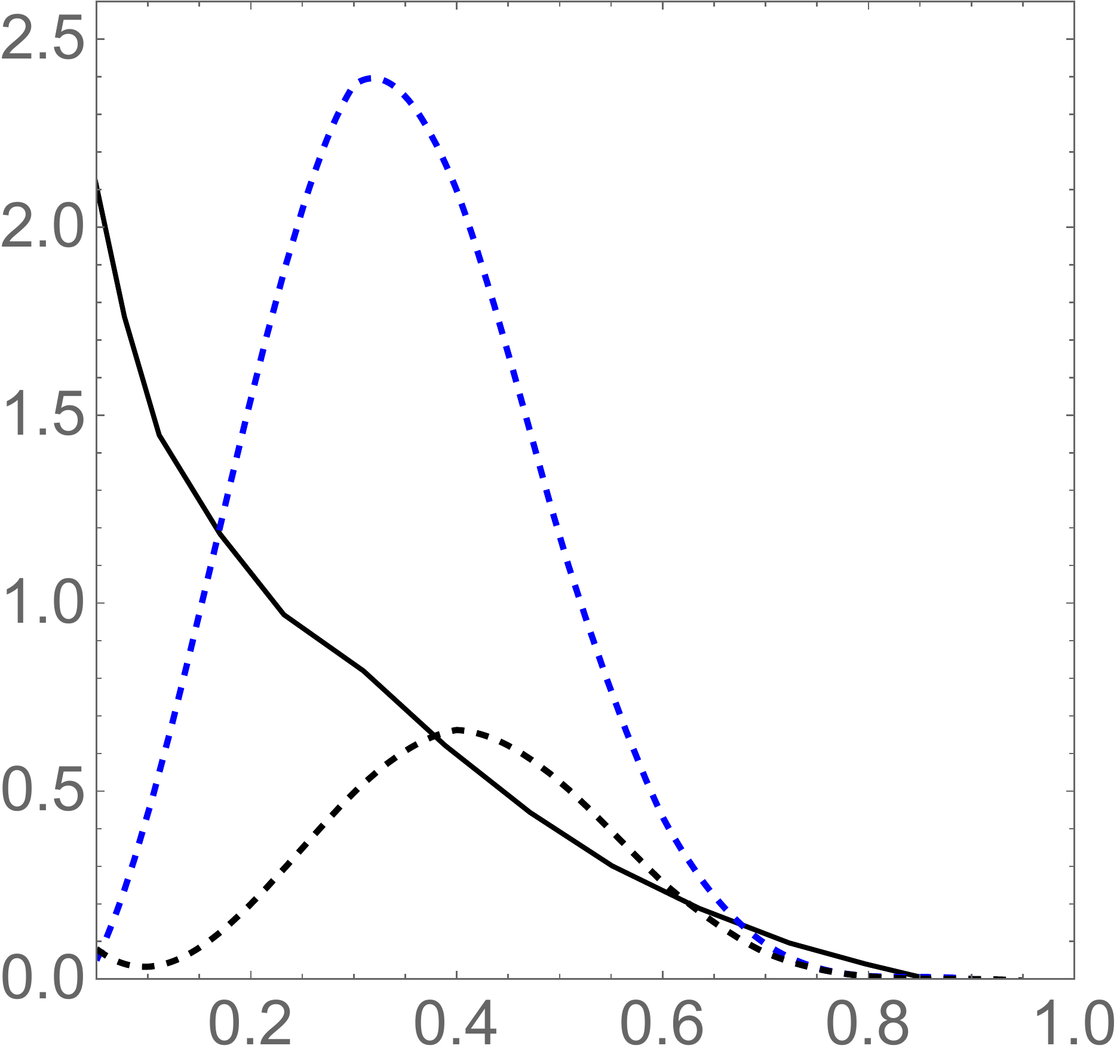}
		\caption{We show the lattice baryonic GPD $H(x,\xi=0,t=-0.69\, {\rm GeV}^2$) versus $x$following from  LaMET in~\protect\cite{2209.05373} (black-solid line),
		and our result (dashed-black line). Our result for  the GPD $H(x,\xi=0,t=0)$ which is the PDF (dashed-blue line), is shown for comparison.}
		\label{fig_H_lat_compared}
	\end{center}
\end{figure}

A more recent lattice study  in~\cite{2209.05373}, extracts the full GPDs from quasi-GPDs using
the lattice momentum effective theory (LaMET)~\cite{Ji:2013dva,Ji:2020ect}. Their result for the
unpolarized nucleon GPD $H(x,\xi,t)$ with symmetric momentum assignment (their Fig.~18),  is shown in 
Fig.~\ref{fig_H_lat_compared} (black-solid line), for zero skewness $\xi=0$ and fixed $t=-0.69\,{\rm GeV}^2$.
Note that we have only show the quark part  of the lattice GPD, and only for $x>0.05$ (the validity of LaMET may even require
a larger lower bound, say $x>0.1$ for current
lattice nucleon  momentum, see discussion in~\cite{Ji:2020ect}).  Note also that the reported lattice data were evolved to the standard normalization scale $\mu^2=4\, GeV^2$.

Our current baryonic LFWFs
are limited tho the 3-quark sector, and therefore to a  low normalization point, where there are no  gluons and $\bar q q$ sea. 
As explained in our previous works, after ``chiral evolution" we used, the ensuing DAs, PDFs, GPDs can be evolved to the chiral normalization
scale $\mu_\chi^2=1\, GeV^2$, and then by DGLAP, to any higher scales, such as  $\mu^2=4\, {\rm GeV}^2$. We have not carried out any of this in the 
present comparison. Our results are also shown in Fig.~\ref{fig_H_lat_compared} for the unpolarized GPD $H(x,\xi=0,t=-0.69\, {\rm GeV}^2)$
(black-dashed line). For comparison, we also show our GPD $H(x,\xi=0,t=0)$ (dashed-blue line), which is effectively the PDF. The agreement between the lattice GPD and 
our GPD for $x>0.4$  is quite reasonable, modulo all the reservations. This is nontrivial, as the underlying $t$ dependence is rather strong, as seen
from comparison to the  $t=0$ GPD curve.

\subsection{Nucleon tomography}
The Fourier transform of the H GPD in the transverse part of the momentum transfer $\Delta_T$ with 
$\Delta_T^2=-t$, yields to a map for the spatial distribution of the partons (mostly  constituent quarks here), with a fixed
longitudinal momentum $x$
\bea
\label{TOMO}
q(x, \xi, b)=\int\frac{d\Delta_\perp}{(2\pi)^2}e^{-i\Delta_\perp\cdot b} H(x, \xi , t=-\Delta_T^2)
\eea
For an estimate, we use the approximate form  (\ref{APPROX}),  to obtain
\bea
\label{TOMO1}
q(x,\xi, b)\sim \frac{\sqrt{1-\xi^2}}{f_s(x)}e^{-\frac{b^2\sqrt{1-\xi^2}}{4f_s(x)}}
\eea
in the DGLAP regime with $x>\xi$.
This impact-parameter-like representation of the GDP in transverse space, provides a 1+2 tomographic
description of the partons inside the nucleon. In general, (\ref{TOMO}-\ref{TOMO1}) allows for the characterization of the 
x-dependence of the parton distribution in a hadron. At large $b$ it reflects on the chiral physics (pion cloud),
while at very low-x it is sensitive to the diffusion of wee partons.

\section{Conclusions}
\label{SEC_CON}
The interest in GPDs stems from their characterization of  the partonic substructure of hadrons,
in terms of parton longitudinal momentum, transverse  position and spin. They can be related to
certain hard processes, allowing their possible extraction from experiment. As such, they are important
tools for the understanding hadronic structures in QCD.

The present work is a modest attempt along these directions, to try to understand how the spatial
distribution of a struck quark in a hadron, depends on its longitudinal momentum $x$. For that, we
made use of some of the results for the LFWFs, we have developed recently~\cite{Shuryak:2022thi,Shuryak:2022wtk}. We recall that 
these LFWFs are not just some parameterizations: they diagonalize certain LF Hamiltonians of increasing complexity, subject to the strictures of
non-perturbative lattice QCD at low resolution (instantons, P-vortices, ...). 

In its simplest form, our LF Hamiltonian  consists of the kinetic and confining contributions
for flavor-symmetric (the $\Delta$-isobar) LFWF, plus a diiquark pairing term between 
flavor-asymmetric
(light $ud$) quarks for the nucleon. So, for the $\Delta$-isobar, the LFWF exhibits triangular
symmetry in the longitudinal parton fractions, while for the nucleon this symmetry is broken by diquark
pairing.

Remarkably, our calculated GPDS for both the nucleon and the $\Delta$-isobar, are found
to be numerically well described by a proposed Gaussian ansatz (\ref{GPD_ansatz}), with
$x,\xi$ dependent width. However,   the width is substantially different from that reported
by other model calculations. 

The GPDs integrated over $x$ give rise to the electromagnetic and gravitational formfactors.
We have used our GPDs with our LFWFs, to recover the electromagnetic fom factors from
our earlier results~\cite{Shuryak:2022wtk}. We also derived anew, the nucleon A-gravitational formfactor.
While both the electromagnetic and gravitational form factors appear similar and dipole-like,
their ratio shows differences, an indication that the charge and mass composition are
spatially distinct.

The GPDs encapsulate a vast amount of dynamical information regarding the 
partonic substructure of hadrons, with fixed $x, \xi, t$ kinematics. However, it is
usually challenging to tie the theoretical insights and results for the GDPs, such 
as the ones based on the LFWFs we have detailed, with their extraction from 
semi-inclusive data. DIS probes the GPDs at
the boundary $x=\xi$ of the finite $x-\xi$-domain, and the convolution integrals 
of the empirical GPDs are limited to the $\xi>0$ region. The sum rules require
integrating over the longitudinal momentum $x$ for fixed $\xi$, while the tomography 
is carried at  $\xi=0$. To relate these separate kinematic regimes, requires the use of
the GDP global analytical properties, and models constrained by these properties, as well as
lattice physics, as we have pursued.

\vskip 1cm
{\bf Acknowledgements}

We thank the members of the QGT collaboration for triggering our motivation for this work.
This work is supported by the Office of Science, U.S. Department of Energy under Contract No. DE-FG-88ER40388.

\appendix

\bibliography{GPDs}
\end{document}